\documentclass[spanish,english,aps,manuscript,aps,preprint]{revtex4}
\usepackage[T1]{fontenc}
\usepackage[latin9]{inputenc}
\usepackage{textcomp}
\usepackage{amsmath}
\usepackage{amssymb}
\usepackage[xdvi]{graphicx}
\usepackage{esint}

\makeatletter
\@ifundefined{textcolor}{}
{%
 \definecolor{BLACK}{gray}{0}
 \definecolor{WHITE}{gray}{1}
 \definecolor{RED}{rgb}{1,0,0}
 \definecolor{GREEN}{rgb}{0,1,0}
 \definecolor{BLUE}{rgb}{0,0,1}
 \definecolor{CYAN}{cmyk}{1,0,0,0}
 \definecolor{MAGENTA}{cmyk}{0,1,0,0}
 \definecolor{YELLOW}{cmyk}{0,0,1,0}
 }

\usepackage{amsfonts}\usepackage{babel}\usepackage{dcolumn}\usepackage{bm}\usepackage{amsfonts}\@ifundefined{textcolor}{}{
 \definecolor{BLACK}{gray}{0}
 \definecolor{WHITE}{gray}{1}
 \definecolor{RED}{rgb}{1,0,0}
 \definecolor{GREEN}{rgb}{0,1,0}
 \definecolor{BLUE}{rgb}{0,0,1}
 \definecolor{CYAN}{cmyk}{1,0,0,0}
 \definecolor{MAGENTA}{cmyk}{0,1,0,0}
 \definecolor{YELLOW}{cmyk}{0,0,1,0}
 }

\usepackage{babel}\addto\shorthandsspanish{\spanishdeactivate{~<>}}

\makeatother

\usepackage{babel}
\addto\shorthandsspanish{\spanishdeactivate{~<>}}

\begin{document}

\title{A study of Wigner functions for discrete-time quantum walks}

\author{M. Hinarejos$^{1}$, M.C. Bañuls$^{2}$ and A. Pérez$^{1}$ }

\affiliation{$\text{\textonesuperior\ }$Departament de Física Teòrica and IFIC,
Universitat de València-CSIC \\
 Dr. Moliner 50, 46100-Burjassot, Spain}

\affiliation{$^{2}$ Max-Planck-Institut für Quantenoptik, Hans-Kopfermann-Str.
1, Garching, D-85748, Germany.}
\begin{abstract}
We perform a systematic study of the discrete time Quantum Walk on
one dimension using Wigner functions, which are generalized to include
the chirality (or coin) degree of freedom. In particular, we analyze
the evolution of the negative volume in phase space, as a function
of time, for different initial states. This negativity can be used
to quantify the degree of departure of the system from a classical
state. We also relate this quantity to the entanglement between the
coin and walker subspaces. 
\end{abstract}
\maketitle

\section{introduction}

Quantum Walks (QW) are considered a as a piece of potential importance
in the design of quantum algorithms \cite{PhysRevA.67.052307,Ambainis2007,ChildsSTOC200359-68,PhysRevA.70.022314,PhysRevA.78.012310},
as it is the case of classical random walks in traditional computer
science. As in the case of random walks, QW's can appear both under
its discrete-time \cite{Aharonov93} or continuous-time \cite{PhysRevA.58.915}
form. Moreover, it has been shown that any quantum algorithm can be
recast under the form of a QW on a certain graph: QWs can be used
for universal quantum computation, this being provable for both the
continuous \cite{PhysRevLett.102.180501} and the discrete version
\cite{PhysRevA.81.042330}. Experiments have been designed or already
performed to implement the QW \cite{Knight2003147,PhysRevA.65.032310,PhysRevA.66.052319,PhysRevA.67.042316,PhysRevA.72.062317,PhysRevA.78.042334,PhysRevA.82.033602,PhysRevLett.104.050502,Xue2009,Zhao2007}.

In this paper, we concentrate on the discrete-time QW on a line. We
perform a systematic study making use of Wigner functions, which are
defined for this problem. Wigner functions \cite{PhysRev.40.749,Hillery1984121}
were introduced as an alternative description of quantum states. They
play an important role in quantum mechanics, having been widely used
in quantum optics to visualize light states. From the experimental
point of view, they provide a way for quantum state reconstruction
via tomography and inverse Radon transformation \cite{citeulike:4313181}.

Wigner functions are quasi-probability distributions in phase space,
meaning that they cannot be interpreted as a probability measure in
momentum \textit{and} space configurations. This is an obvious fact
for any quantum description, and only marginal distributions can be
associated to probabilities in position \textit{or} momentum (or any
linear combination, i.e. any \textit{quadrature}). In fact, Wigner
functions can take negative values, thus invalidating a direct link
to a probability distribution. This caveat, however, turns out to
be a potential advantage, for it can be used to identify {}``true''
quantum states. More precisely, the volume of the negative part of
the Wigner function, its \textit{negativity}, has been suggested as
a figure of merit to quantify the degree of \textit{non-classicality}
\cite{1464-4266-6-10-003}. This idea has been recently exploited
\cite{PhysRevLett.106.010403} to directly estimating nonclassicality
of a state by measuring its distance from the closest one with a positive
Wigner function.

When dealing with the discrete QW, one has to account for the extra
degree of freedom (in addition to the spatial motion): the coin. We
consider the simplest case of a two level coin. Therefore, the Wigner
function has to incorporate this extra index and, with the prescription
we use, it turns into a matrix.  We will propose a rather straightforward
extended definition of negativity for this Wigner {}``function''.
Then the question arises, what kind of states of the QW are nonclassical?
Does this \textit{quantumness} increase in time, as the QW evolves
through its unitary evolution? We want to explore these questions
using the Wigner function. A different topic, although it is closely
related to the previous one, is whether this nonclassicality can be
related to the entanglement between the walker and the coin, since
this quantity is also evolving during the QW evolution. 

This paper is organized as follows. In Sect. II we review the main
definitions pertinent to the QW on a line. Sect. III introduces the
Wigner function for our problem and the main associated properties.
We present some examples showing our numerical results for the Wigner
function evolution in Sect. IV. In Sect. V we define an extension
of the negativity to the QW, based on the proposal made in \cite{1464-4266-6-10-003}
for a scalar function. We end in Sect. VI by summarizing our main
results and conclusions.

\section{Discrete-time QW walk on a lattice}

The discrete-time QW on a line is defined as the evolution of a one-dimensional
quantum system following a direction which depends on an additional
degree of freedom, the coin (or chirality), with two possible states:
{}``left\textquotedblright{}\ $|L\rangle$\ or {}``right\textquotedblright{}\ $|R\rangle$.
The total Hilbert space of the system is the tensor product $H_{s}\otimes H_{c}$,
where $H_{s}$ is the Hilbert space associated to the lattice, and
$H_{c}$ is the coin Hilbert space. Let us call $T_{-}$ ($T_{+}$)
the operators in $H_{s}$ that move the walker one site to the left
(right), and $|L\rangle\langle L|$ , $|R\rangle\langle R|$ the chirality
projector operators in $H_{c}$. The QW is defined by 
\begin{equation}
U(\theta)=T_{-}\otimes|L\rangle\langle L|\text{ }C(\theta)+T_{+}\otimes|R\rangle\langle R|\text{ }C(\theta),\label{Ugen}
\end{equation}
 where $C(\theta)=\sigma_{z}\cos\theta+\sigma_{x}\sin\theta$, and
$\sigma_{z}$, $\sigma_{x}$ are Pauli matrices acting on $H_{c}$.
For $\theta=\pi/4$ the operator $C(\theta)$ becomes the Hadamard
transformation. The unitary operator $U(\theta)$ transforms the state
in one time step as 
\begin{equation}
|\psi(t+1)\rangle=U(\theta)|\psi(t)\rangle.\label{evolution}
\end{equation}
 The state at time $t$ can be expressed as the spinor 
\begin{equation}
|\psi(t)\rangle=\sum\limits _{n=-\infty}^{\infty}\left[\begin{array}{c}
a_{n}(t)\\
b_{n}(t)
\end{array}\right]|n\rangle,\label{spinor}
\end{equation}
 where the upper (lower) component is associated to the right (left)
chirality, and $\left\{ \mid n\rangle/n\in\mathbb{Z}\right\} $ is
a basis of position states on the lattice. A basis in the whole Hilbert
space can be constructed as the set of states $\left\{ \mid n,\alpha\rangle=\mid n\rangle\otimes\mid\alpha\rangle/n\in\mathbb{Z};\alpha=L,R\right\} $.

\section{Wigner functions for the quantum walk}

An important tool in some fields related to quantum physics is the
use of quasi-probability distributions. Wigner functions constitute
the major example, although other functions as the Glauber-Sudarshan
$P$ function \cite{PhysRev.131.2766,PhysRevLett.10.277} or the Husimi
$Q$ function \cite{hu40} are commonly used in quantum optics. For
the case of a one-dimensional system with continuous position $x$
and conjugate momentum $p$, the Wigner function is defined as \cite{PhysRev.40.749}:

\begin{equation}
W(x,p)=\frac{1}{\pi}\intop_{-\infty}^{\infty}\psi^{*}(x-y)\psi(x+y)e^{-2ipy}dy,\label{Wignercontin}
\end{equation}

where $\psi(x)$ is the wave function for the system in a state $|\psi\rangle$,
and we are using units such that $\hbar=1$. A number of properties
can be derived from the definition, the most important ones giving
the probability in position (momentum) space as the marginal distributions
obtained by integration over momentum (position), respectively. We
refer the interested reader to references \cite{Hillery1984121,citeulike:4313181}
for an overview about these properties.

We are interested in describing the QW with the help of Wigner functions.
The case of a finite lattice, with periodic conditions, has been widely
studied \cite{Wootters1987,PhysRevA.70.062101,Lopez2003,PhysRevA.68.052305}
. Here we want to describe the QW on an infinite lattice of equally-spaced
positions. A proposal for the Wigner function to study this problem
will be published elsewhere \cite{HBP2012}. As discussed in this
reference, due to the discreteness of the phase space, one needs to
double the phase space in order to fulfill the necessary properties
of the Wigner function. This doubling feature is a characteristic
of discrete Wigner functions \cite{1751-8121-44-34-345305,Lopez2003},
and we will return to this point later.

Secondly, we need to incorporate the chiral (or spin) degree of freedom.
To do this, we consider the walker as a spin 1/2 particle moving on
the lattice. This analogy allows us to make the connection to the
extensive literature of Wigner functions describing particles with
spin. They have been extended to relativistic particles \cite{PhysRevD.13.950}
and widely used in kinetic theory \cite{nla.cat-vn1394639}, nuclear
physics \cite{DiazAlonso:1991hy} or to describe neutrino propagation
in matter \cite{Sirera:1998ia}. Within this approach, wave functions
as defined in \cite{HBP2012} are to be replaced by (Dirac) spinors.
Since we are interested in a nonrelativistic description, we simply
use 2-dimensional spinors and define:

\begin{equation}
W_{\alpha\beta}(n,k,t)=\frac{1}{\pi}e^{ikn}\sum_{l\in\mathbb{Z}}\langle l,\alpha\mid\psi(t)\rangle\langle\psi(t)\mid n-l,\beta\rangle e^{-2ikl}.\label{Wignerdiscrete}
\end{equation}

In the latter equation, $|\psi(t)\rangle$ represents the state of
the QW at time $t$. This definition can be extended to the case when
the state of the system (walker plus coin) is described by the density
matrix $\rho(t).$ In such a case, we would have 
\begin{equation}
W_{\alpha\beta}(n,k,t)=\frac{1}{\pi}e^{ikn}\sum_{l\in\mathbb{Z}}\langle l,\alpha\mid\rho(t)\mid n-l,\beta\rangle e^{-2ikl}.
\end{equation}

In what follows, we will omit the chirality subindices, so that $W(n,k,t)$
will represent an hermitian matrix in chiral space. Also, it will
be understood that summations with natural indices are performed over
all integers in $\mathbb{Z}$. Substitution of Eq. (\ref{spinor})
into Eq. (\ref{Wignerdiscrete}) immediately gives

\begin{eqnarray}
W(n,k,t) & = & \frac{1}{\pi}e^{ikn}\sum_{l}e^{-2ikl}\left(\begin{array}{cc}
a_{l}(t)a_{n-l}^{*}(t) & a_{l}(t)b_{n-l}^{*}(t)\\
b_{l}(t)a_{n-l}^{*}(t) & b_{l}(t)b_{n-l}^{*}(t)
\end{array}\right).
\end{eqnarray}

Notice that the Wigner function is defined over the phase space associated
to the Hilbert space of the lattice. This phase space is given by
pairs $(n,k)$, with $n\in\mathbb{Z}$ and $k\in[-\pi,\pi[$. These
coordinates are not to be confused with the position and momentum
of the real system, although they are associated to them. To avoid
ambiguity, we will refer to $n$ and $k$ as the phase space spatial
and momentum coordinates. 

Using the above definitions, there are many properties that can be
proven in a straightforward way. The most important ones are given
below.

By performing the integration over $k$ one readily obtains, for even
values of the spatial phase space coordinate,

\begin{equation}
\intop_{-\pi}^{\pi}W(2n,k,t)dk=2\left(\begin{array}{cc}
\mid a_{n}(t)\mid^{2} & a_{n}(t)b_{n}^{*}(t)\\
b_{n}(t)a_{n}^{*}(t) & \mid b_{n}(t)\mid^{2}
\end{array}\right),
\end{equation}
 while, for odd values 
\begin{equation}
\intop_{-\pi}^{\pi}W(2n+1,k,t)dk=0.
\end{equation}

This result is a consequence of the above mentioned doubling feature.
From here, the spatial probability distribution can be recovered by
performing the trace over coin variables:

\begin{equation}
\frac{1}{2}Tr\left(\intop_{-\pi}^{\pi}W(2n,k,t)dk\right)=\mid a_{n}(t)\mid^{2}+\mid b_{n}(t)\mid^{2}=P(n,t),
\end{equation}

where $P(n,t)$ stands for the probability of detecting the walker
at position $n$, regardless of the coin state.

To obtain the distribution in momentum space we start from Eq. (\ref{spinor})
and introduce the quasi-momentum basis $\left\{ \mid k\rangle/k\in[-\pi,\pi[\right\} $
(restricted to the first Brillouin zone), which is related to the
spatial basis $\left\{ \mid n\rangle/n\in\mathbb{Z}\right\} $ via
Fourier discrete transformation, i.e.:

\begin{equation}
\langle n\mid k\rangle=\frac{1}{\sqrt{2\pi}}e^{ink}.
\end{equation}
 After projecting over a given $|k\rangle$ one obtains

\begin{equation}
\langle k\mid\psi(t)\rangle=\left(\begin{array}{c}
\tilde{a}_{k}(t)\\
\tilde{b}_{k}(t)
\end{array}\right),\label{pureink}
\end{equation}
 where

\begin{equation}
\tilde{a}_{k}(t)\equiv\frac{1}{\sqrt{2\pi}}\sum_{n}e^{-ink}a_{n}(t);\,\,\,\,\tilde{b}_{k}(t)\equiv\frac{1}{\sqrt{2\pi}}\sum_{n}e^{-ink}b_{n}(t)
\end{equation}
 are the chirality components of the wave function in momentum space.
By introducing the closure relation for the basis of states $\{\mid k\rangle\}$,
one can relate the Wigner function to the matrix elements of $\rho(t)$
in momentum space: 
\begin{equation}
W_{\alpha\beta}(n,k,t)=\frac{1}{\pi}\int_{-\pi}^{\pi}e^{in(q-k)}\langle q,\alpha\mid\rho(t)\mid2k-q,\beta\rangle dq,
\end{equation}

with $\mid k,\alpha\rangle=\mid k\rangle\otimes\mid\alpha\rangle$.
Summation over $n$ leads to

\begin{equation}
M(k,t)\equiv\sum_{n}W(n,k,t)=2\langle k\mid\rho\mid k\rangle,\label{matrixm}
\end{equation}
 where use was made of the equation $\sum_{n}e^{ink}=2\pi\delta(k,2\pi)$,
with $\delta(k,2\pi)\equiv\sum_{m}\delta(k+2\pi m)$ the {}``Dirac
comb'' function. Obviously, $M(k,t)$ is also a $2\times2$ matrix.
For a pure state we have, using Eq. (\ref{pureink}),
\begin{equation}
M(k,t)=2\langle k\mid\psi(t)\rangle\langle\psi(t)\mid k\rangle=2\left(\begin{array}{cc}
\mid\tilde{a}_{k}(t)\mid^{2} & \tilde{a}_{k}(t)\tilde{b}_{k}^{*}(t)\\
\tilde{b}_{k}(t)\tilde{a}_{k}^{*}(t) & \mid\tilde{b}_{k}(t)\mid^{2}
\end{array}\right).
\end{equation}

The diagonal components of this matrix give (up to a factor 2) the
probability in momentum space, when chirality is specified, whereas
non-diagonal components correspond to coherences between different
chiralities. Now, the trace over chirality provides the probability
in momentum space, when chirality is not measured:

\begin{equation}
\frac{1}{2}Tr\{M(k,t))\}=\tilde{a}_{k}^{2}(t)+\tilde{b}_{k}^{2}(t)\equiv P(k).
\end{equation}

Since the evolution in momentum space is diagonal in $k$ and given
by a unitary transformation (see, for example \cite{Ambainis2001}),
it can be easily proven that the magnitude $P(k)$ defined above remains
constant with time.

Another important property that can be derived for the Wigner function
of the QW is a recursion formula relating $W(n,k,t+1)$ to other components
of this function at time $t$. Using Eq. (\ref{evolution}) one obtains,
after some algebra:

\selectlanguage{spanish}%
\begin{align}
W(n,k,t+1) & =M_{R}W(n-2,k,t)M_{R}^{\dagger}+e^{-2ik}M_{R}W(n,k,t)M_{L}^{\dagger}\nonumber \\
+ & e^{2ik}M_{L}W(n,k,t)M_{R}^{\dagger}+M_{L}W(n+2,k,t)M_{L}^{\dagger},\label{recWigner}
\end{align}

\selectlanguage{english}%
where $M_{L}=(|L\rangle\langle L|)C(\theta)$ and $M_{L}=(|R\rangle\langle R|)C(\theta)$.
An immediate consequence of this recursion formula is that sites with
even $n$ evolve independently of those with odd $n$.

\section{Numerical results}

\begin{figure}
\includegraphics[clip,width=12cm]{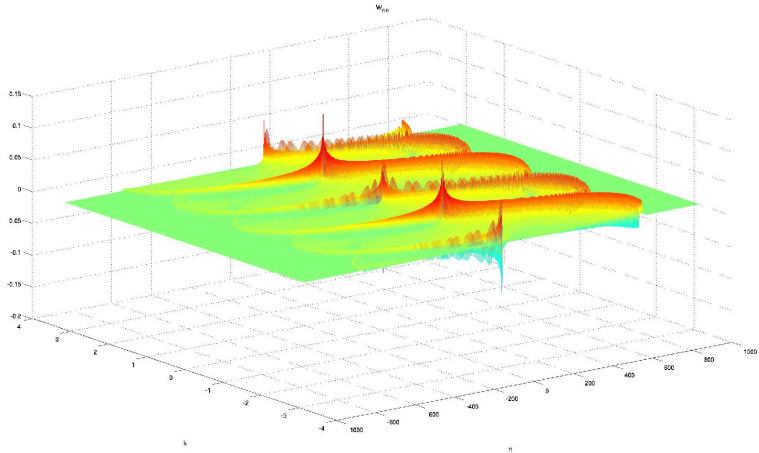}

\caption{(Color online): $W_{RR}$ component of the Wigner function with initial
conditions given as in (\ref{initial_loc}), after 500 iterations.}

\label{WRR_loc} 
\end{figure}

In what follows, we will discuss a couple of examples showing the
main features of the Wigner function. We have numerically simulated
the QW evolution for various initial states, and explicitly computed
the Wigner function. We take the lattice large enough, so that boundaries
do not need to be considered. In practice, this is equivalent to assuming
an infinite lattice. In the cases we will consider here, the initial
state is such that at $t=0$ the Wigner function is non-vanishing
only at phase space points with even $n$. Then, Eq. (\ref{recWigner})
warrants that $W(2s+1,k,t)=0$, $\forall s\in\mathbb{Z}$, at any
time. Therefore, we only plot the Wigner function over the part of
the phase space with even spatial coordinate. Fig \ref{WRR_loc} shows
$W_{RR}(n,k,t)$ for $t=500$, with initially localized conditions,

\selectlanguage{spanish}%
\begin{equation}
\mid\psi(0)\rangle=\frac{1}{\sqrt{2}}(\mid R\rangle+i\mid L\rangle)\otimes\mid0\rangle\label{initial_loc}
\end{equation}

\selectlanguage{english}%
The initial Wigner function for this state can be easily evaluated,
with the result\foreignlanguage{spanish}{
\begin{equation}
W(n,k,0)=\frac{1}{2\pi}\delta_{n,0}\left(\begin{array}{cc}
1 & -i\\
i & 1
\end{array}\right).
\end{equation}
}

\begin{figure}
\begin{minipage}[t]{1\columnwidth}%
\includegraphics[width=5.5cm]{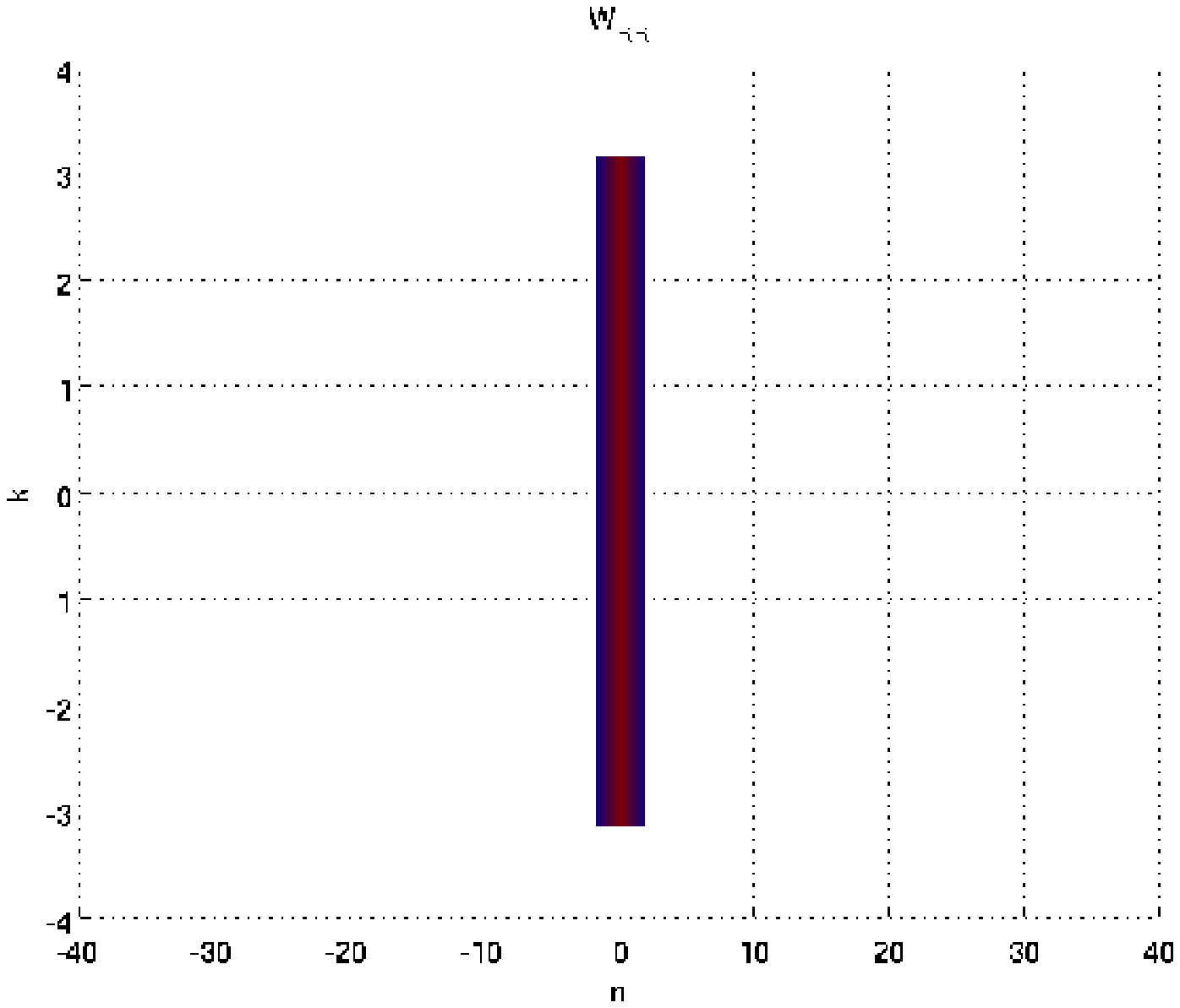}\includegraphics[width=5.5cm]{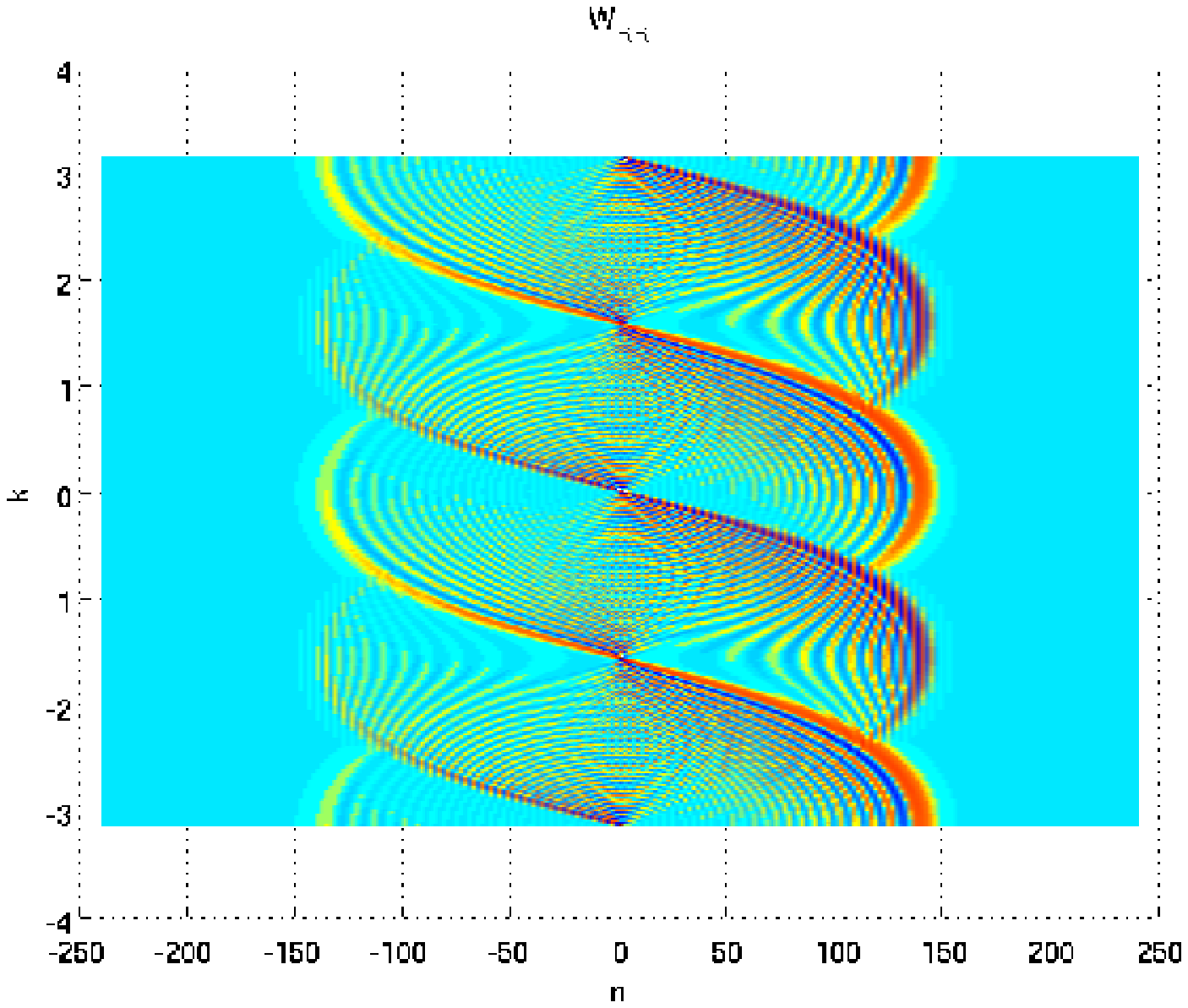}\includegraphics[width=5.5cm]{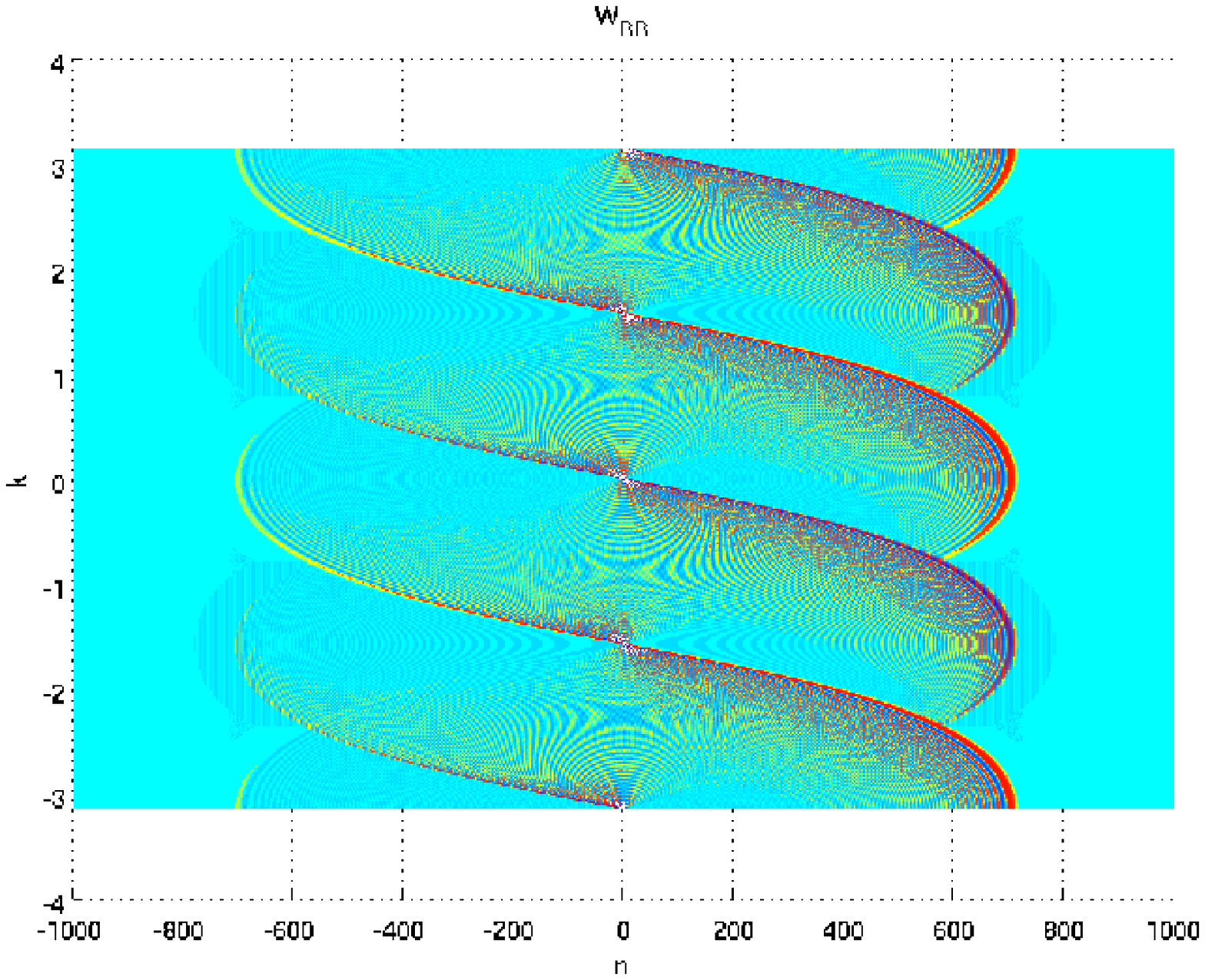}%
\end{minipage}

\caption{(Color online): Contour plots showing the time evolution of the $W_{RR}$
component of the Wigner function starting from the localized state
(\ref{initial_loc}). From left to right, the sub figures correspond
to $t=0$, $t=100$ and $t=500$, respectively. }

\label{WRR_loc_top}
\end{figure}

\begin{figure}
\includegraphics[width=12cm]{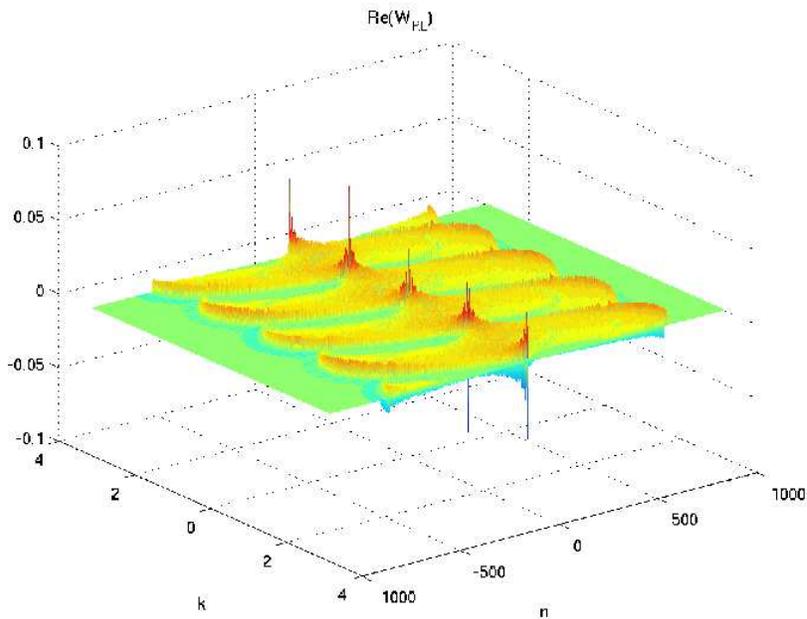}\caption{(Color online): Real part of the $W_{RL}$ component of the Wigner
function with initial conditions given as in (\ref{initial_loc}),
after 500 iterations.}

\label{ReWRL_loc}
\end{figure}

Since the matrix defining the Wigner function is Hermitian, this component
has no imaginary part. The evolved component $W_{RR}$ after 500 iterations
is shown in Fig. \ref{WRR_loc}, and the time evolution of this component
can be seen on Fig. \ref{WRR_loc_top}. One can observe an intricate
structure, arising from interference effects. Notice, for example,
the similarity with the threads mentioned in \cite{Lopez2003}. It
is interesting to mention that, although the Wigner function expands
in space, as the walker distribution broadens, it keeps the same structure.
The rest of components of the Wigner function show a similar appearance.
As an example, we have represented in Fig. \ref{ReWRL_loc} the real
part of the off-diagonal component $W_{RL}$ for $t=500$, starting
from the localized state Eq. (\ref{initial_loc}).

\begin{figure}
\includegraphics[width=10cm]{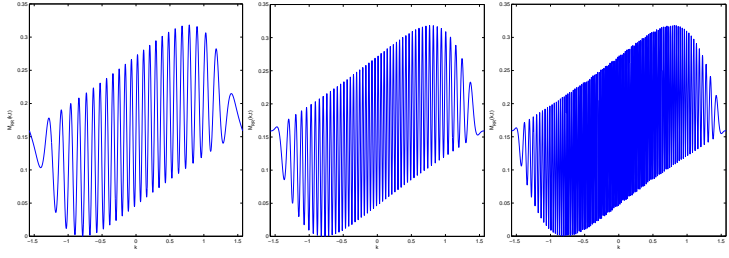}\caption{$M_{RR}(k,t)$ component with the same initial conditions as before,
corresponding to times $t=50$, $t=100$ and $t=200$ (left to right).}

\label{momentumRR}
\end{figure}

The momentum distribution is obtained from Eq. (\ref{matrixm}). We
show in Fig. \ref{momentumRR}, as an example, the RR component of
$M(k,t)$ plotted for $t=50,100,200$ and the same initial condition
(\ref{initial_loc}) as before. Since the moment $k$ is bounded,
the distribution becomes more intricate as the QW evolves. At later
times, the figure shows more oscillations, although the envelop remains
constant, in accordance to the self-resemblance of the Wigner function
as time increases.

For comparison, we investigate the Wigner function for a different
initial condition. It corresponds to a {}``Schrödinger cat'' in
two positions $|\pm a\rangle$ which are entangled with two chiralities,
i.e.:

\begin{equation}
\mid\psi(0)\rangle=\frac{1}{\sqrt{2}}(\mid a,R\rangle+i\mid-a,L\rangle).\label{initial_cat}
\end{equation}

The initial Wigner function for this state is now
\begin{equation}
W(n,k,0)=\frac{1}{2\pi}\left(\begin{array}{cc}
\delta_{n,2a} & -ie^{-2ika}\delta_{n,0}\\
ie^{2ika}\delta_{n,0} & \delta_{n,-2a}
\end{array}\right).
\end{equation}

\begin{figure}
\includegraphics[width=12cm]{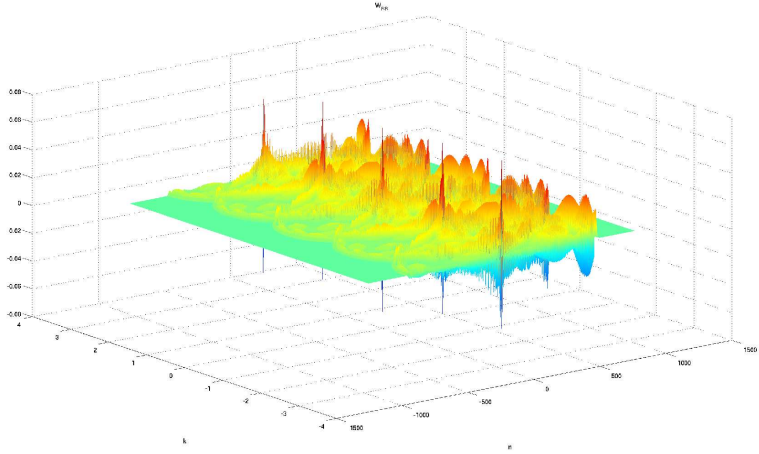}

\caption{(Color online): $W_{RR}$ component of the Wigner function with initial
conditions given as in (\ref{initial_cat}), and $a=10$, after 500
iterations.}
\label{WRR_schro}
\end{figure}

\begin{figure}
\includegraphics[width=5.5cm]{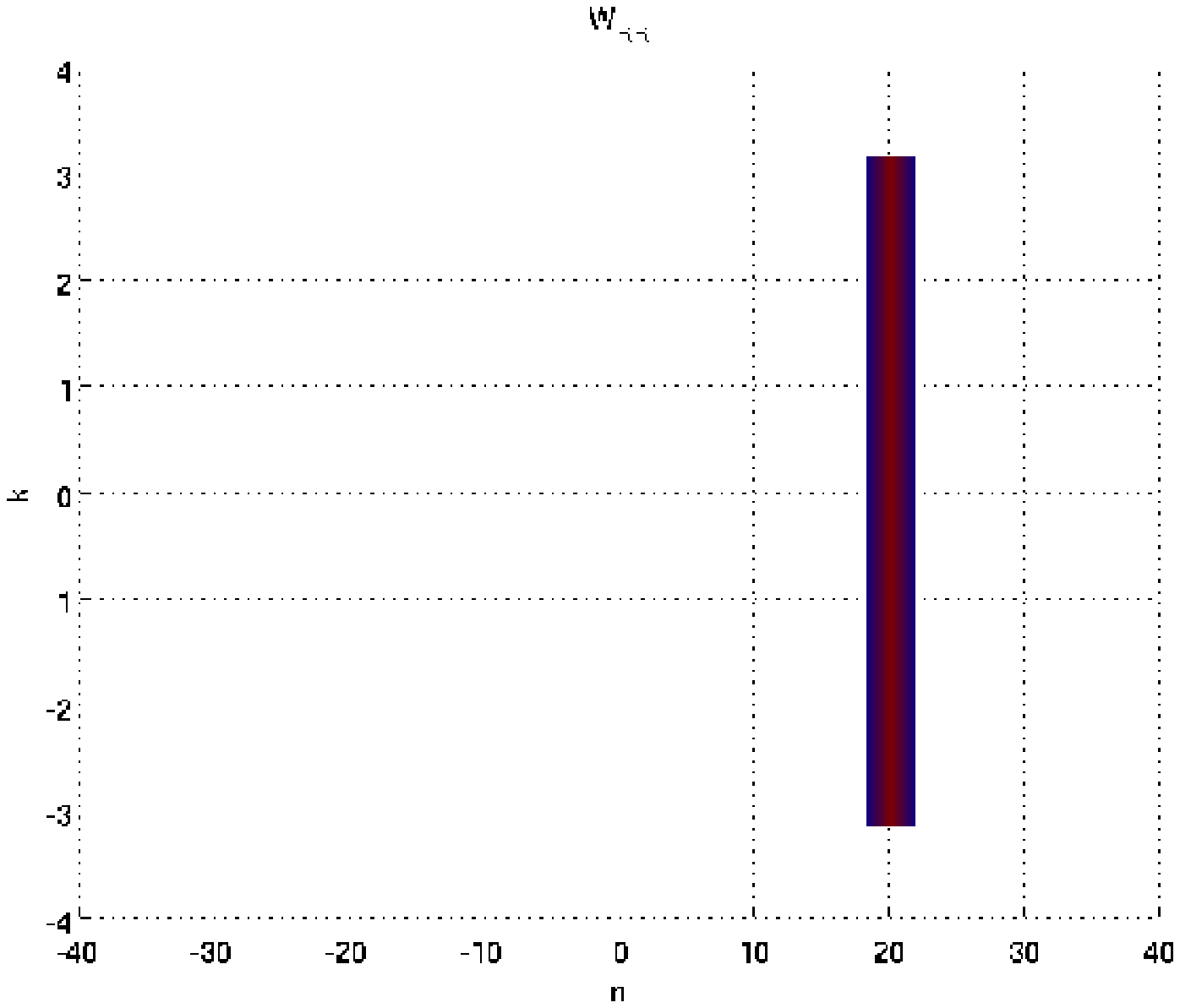}\includegraphics[width=5.5cm]{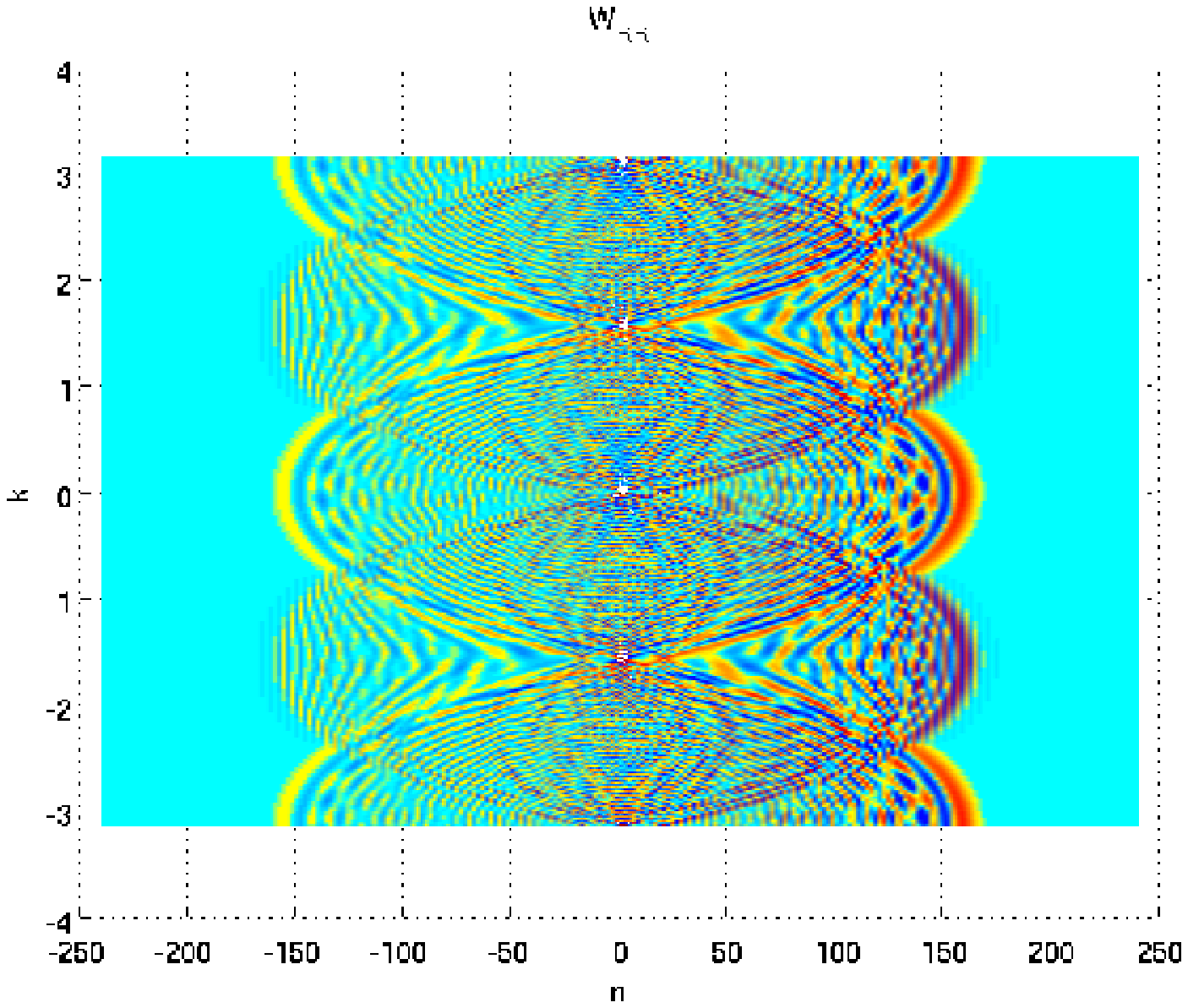}\includegraphics[width=5.5cm]{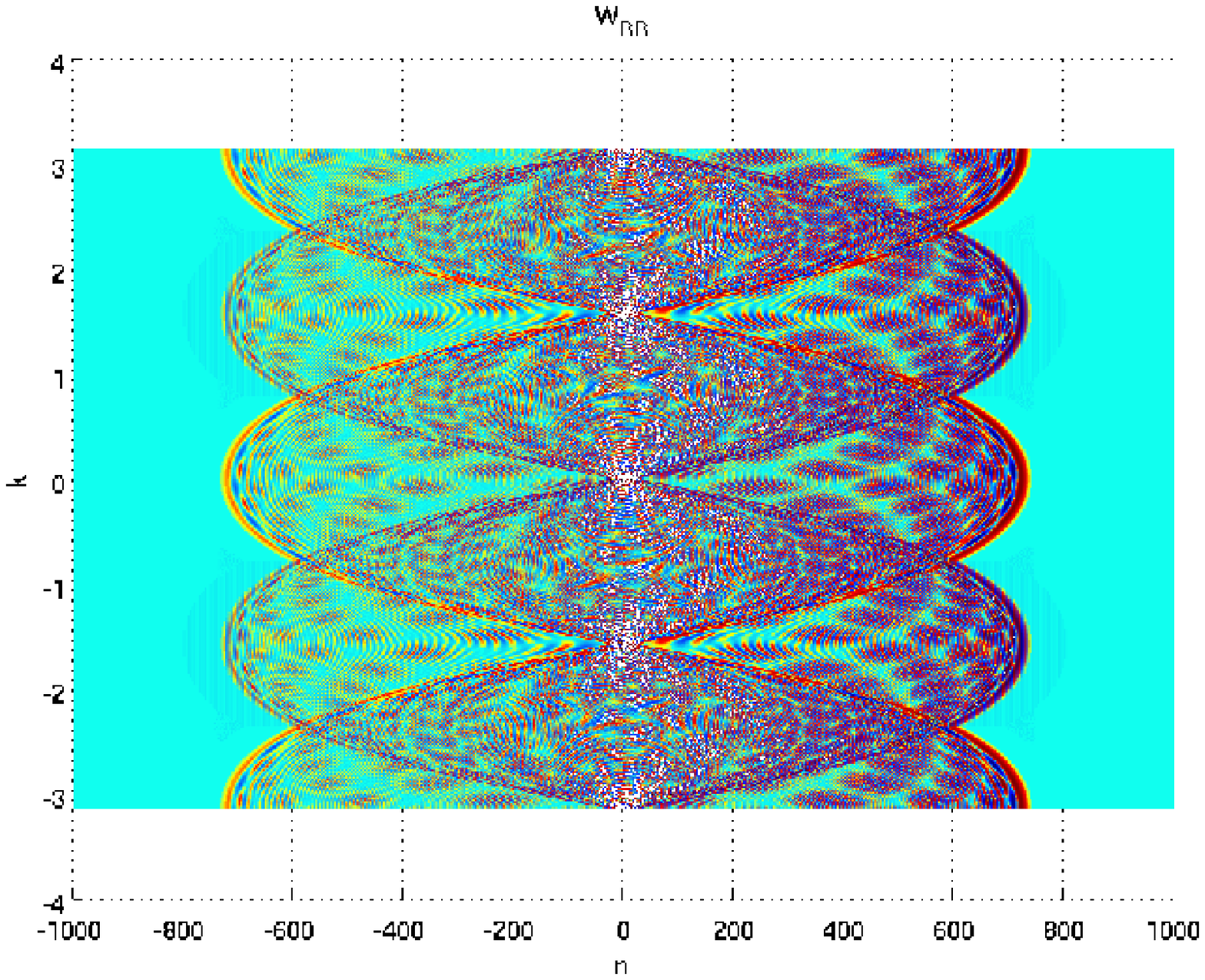}

\caption{(Color online): Contour plots showing the time evolution of the $W_{RR}$
component of the Wigner function starting from the localized state
(\ref{initial_cat}) with $a=10$. From left to right, the sub figures
correspond to $t=0$, $t=100$ and $t=500$. }

\label{WRR_schro_top}
\end{figure}

Notice that $a=0$ reproduces the localized state described above.
The evolved component $W_{RR}$ of the Wigner function is shown in
Fig. \ref{WRR_schro}. Fig. \ref{WRR_schro_top} reveals the time
evolution of this component. As compared to the previous case, it
shows an even more complicated structure, thus suggesting a less classical
state. This suggestion will be confirmed in the next section by comparing
the negativity for both cases. 


\section{Negativity of the wigner function in the quantum walk}

Consider the Wigner function, as defined in Eq. (\ref{Wignercontin}).
As already mentioned, the fact that this function can take negative
values implies that it cannot be considered as a classical probability
distribution. Therefore, non positivity of the Wigner function can
be interpreted as a measure of the non-classicality of the system.
In quantum optics, this is interpreted as a signature for non-classical
states of light, caused by a quantum interference phenomenon. In the
context of finite dimensional systems, this idea is exploited in \foreignlanguage{spanish}{\cite{Franco2006}}
to establish a criterion for entanglement in a system of two spin
1/2 particles. In \foreignlanguage{spanish}{\cite{PhysRevA.74.042323}},
a connection is found between entanglement and the negativity of the
Wigner function for hyperradial s-waves. Such states, in $D=2d$ dimensions,
can be interpreted as the wave function of two entangled particles
in $d$ dimensions.

A quantitative measure of non-classicality is given by the negativity,
as defined in \cite{1464-4266-6-10-003}. In the continuous case,
with variables $x$ and $p$, this volume can be written as:

\begin{equation}
\delta(W)=\int_{-\infty}^{\infty}\int_{-\infty}^{\infty}[\mid W(x,p)\mid-W(x,p)]dpdx=\int_{-\infty}^{\infty}\int_{-\infty}^{\infty}\mid W(x,p)\mid dpdx-1.\label{negscalar}
\end{equation}

In deriving the latter equality, we made use of the fact that the
total probability is normalized to one, so that $\int_{-\infty}^{\infty}\int_{-\infty}^{\infty}W(x,p)dpdx=1$.
We will use $\delta(W)$, adopted to the discrete case, to compute
the negativity of the Wigner function, and to explore whether we can
relate the deviations from a classical behavior of the QW to the entanglement
between the walker and the coin.

In our case, the Wigner function is a 2x2 matrix, and variables $x$,
$p$ are to be replaced by $n$ and $k$, as defined in Sect. III.
We propose, as a possible generalization of the above equation \cite{HBP2012}:

\begin{equation}
\delta(W)=\sum_{n}\int_{-\pi/2}^{\pi/2}[\mid\mid W(n,k)\mid\mid-Tr(W(n,k))]dk=\sum_{n}\int_{-\pi/2}^{\pi/2}\mid\mid W(n,k)\mid\mid dk-1,\label{negativity}
\end{equation}

and where we omitted the dependence on time to simplify the notation.
Again, the latter equality arises as a consequence of normalization.
As a measure of the norm $||A||$ of a matrix $A$, we adopt the trace
norm, defined as:

\begin{equation}
\mid\mid A\mid\mid\equiv Tr\sqrt{A^{\dagger}A},
\end{equation}

where $A^{\dagger}$ denotes the hermitian conjugate of $A$. If $\lambda_{1}(n,k)$,
$\lambda_{2}(n,k)$ are the eigenvalues of $W(n,k)$ for a given $n$
and $k$, one obtains $||W||=\mid\lambda_{1}(n,k)\mid+\mid\lambda_{2}(n,k)\mid$.
In this way, Eq. (\ref{negativity}) adopts the following form:

\begin{equation}
\delta(W)=\int_{-\pi/2}^{\pi/2}\sum_{n}\left(\mid\lambda_{1}(n,k)\mid+\mid\lambda_{2}(n,k)\mid-\lambda_{1}(n,k)-\lambda_{2}(n,k)\right)dk,
\end{equation}

which can be regarded as a natural generalization of (\ref{negscalar}).

\begin{figure}
\includegraphics[width=8cm]{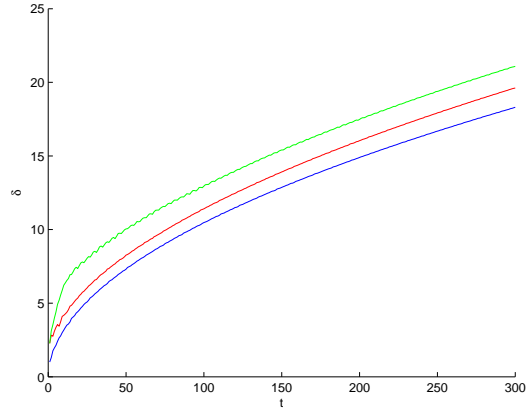}\caption{(Color online): Negativity, as a function of time, for the localized
state $a=0$ (blue, bottom curve) and for the Schrödinger cat state
with $a=4$ (red, middle curve) and $a=30$ (green, top curve).}

\label{fignegativity} 
\end{figure}

We have numerically calculated the negativity (\ref{negativity})
as a function of time, for the same initial conditions considered
in the previous section: the localized state and two different cat
states (Fig. \ref{fignegativity}).

We immediately observe that one obtains a higher degree of {}``quantumness''
for the Schrödinger cat states, as compared to the localized state.
This was expected from the higher degree of complexity observed in
the Wigner function for the cat state, and implies a larger amount
of interference effects. Higher values of the initial separation yield
a larger negativity.

The question arises whether this higher degree of quantumness will
also imply a higher entanglement between the walker and the coin.
To this end, we use the entropy of entanglement as a quantity to characterize
this property. To be more precise, we compute the quantity

\begin{equation}
S(t)=-Tr\{\rho_{c}(t)\log_{2}\rho_{c}(t)\},
\end{equation}
 where $\rho_{c}(t)=Tr_{S}\{\rho(t)\}$ is the density matrix for
the coin, which is obtained by tracing out the spatial degrees of
freedom.

\begin{figure}
\includegraphics[width=8cm]{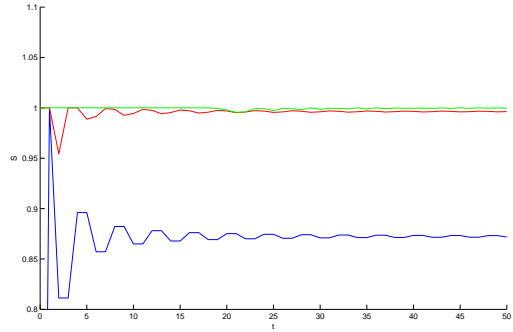}

\caption{Entropy of the localized state $a=0$ (blue, bottom curve) and for
the Schrödinger cat state with $a=4$ (red, middle curve) and $a=30$
(green, top curve).}
\label{figentrop}
\end{figure}

Fig. \ref{figentrop} shows $S(t)$ for the localized state ($a=0$)
and for the cat state with different values of the (half) separation
$a$. The entanglement entropy is bounded by the (log of the) dimension
of the coin space: $S(t)\le1$, and we can see that the time scale
to reach this maximum is shorter for larger values of $a$, which
also correspond to higher degrees of negativity, as seen in previous
figures. 

\section{Conclusions}

In this paper, we have studied the discrete-time quantum walk on a
one-dimensional lattice using the Wigner function. We have explored
the potential of this phase-space representation to characterize the
dynamics of the system. Differently to the case of a scalar wave function,
we now have a $2\times2$ matrix, defined over a phase space $(n,k)$.
We have examined two cases, which correspond to different initial
conditions. The first case starts with an initially localized state
at $n=0$, as considered in many works in the literature. Its Wigner
function shows a quite intricate structure, which is build up by interference
effects on the lattice. We have also considered Schrödinger cat states,
in which space and coin are initially entangled. It is apparent, from
the Wigner function plots, that such states are capable of building
up higher interference effects, and thus describe states that are
even {}``less classical'' than the previous one. In order to quantify
this effect, we have computed the \textit{negativity} associated to
the Wigner function \cite{1464-4266-6-10-003}, as defined by the
negative volume of the function in phase space. This definition has
been extended to our case by the use of the trace norm of the Wigner
matrix. As expected, the cat states give rise to a larger extent of
negativity, which is even larger when the initial separation of the
cat state is increased. In accordance to these ideas, one also obtains
that the coin-walker entanglement evolves faster for the latter states.
Altogether, the time evolution of the QW translates into an evolving
state which separates from classicality. This separation, as time
goes on, was expected from the fact that the walker distribution expands
faster (with a quadratic deviation $\sigma\sim t$ ) than its classical
counterpart (the random walk, for which $\sigma\sim\sqrt{t}$).
\begin{acknowledgments}
This work was supported by the Spanish Grants FPA2011-23897, 'Generalitat
Valenciana' grant PROMETEO/2009/128, and by the DFG through the SFB
631, and the Forschergruppe 635.
\end{acknowledgments}
\bibliographystyle{apsrev}
\bibliography{qwalks,wigner}

\end{document}